\documentclass[twocolumn]{article}

\usepackage{arxiv}

\usepackage[all]{nowidow}

\usepackage[utf8]{inputenc} 
\usepackage[T1]{fontenc}    
\usepackage{hyperref}       
\usepackage{url}            
\usepackage{booktabs}       
\usepackage{amsfonts}       
\usepackage{nicefrac}       
\usepackage{microtype}      

\usepackage{float}  
\usepackage{graphicx}
\usepackage[labelfont=bf]{caption}
\usepackage{subcaption}     
\usepackage{tcolorbox}      

\title{Sanguine: Visual Analysis for Patient Blood Management}

\def \shortauthor {Lin et al.}
\author{
  Haihan Lin$^*$ \\
  University of Utah
     \And
   Ryan A. Metcalf$^*$ \\
   University of Utah 
      \And
   Jack Wilburn \\
   University of Utah 
      \And
      Alexander Lex \\
  University of Utah \\
  \texttt{alex@sci.utah.edu}
}

\begin{document}
\twocolumn[
  \begin{@twocolumnfalse}
    \maketitle
    
\begin{abstract}
Blood transfusion is a frequently performed medical procedure in surgical and nonsurgical contexts. Although it is often necessary or even life-saving, it has been identified as one of the most overused procedures in hospitals. Unnecessary transfusions not only waste resources but can also be detrimental to patient outcomes. Patient blood management (PBM) is the clinical practice of optimizing transfusions and associated outcomes. In this paper, we introduce Sanguine, a visual analysis tool for transfusion data and related patient medical records. Sanguine was designed with two user groups in mind: PBM experts who oversee blood management practices across an institution and clinicians performing transfusions. PBM experts use Sanguine to explore and analyze transfusion practices and its associated medical outcomes. They can compare individual surgeons, or compare outcomes or time periods, such as before and after an intervention regarding transfusion practices. PBM experts then curate and annotate views for communication with clinicians, with the goal of improving their transfusion practices. We validate the utility and effectiveness of Sanguine through case studies. 
\end{abstract}
 
\vspace{2mm}    
\keywords{Data Visualization\and Patient Blood Management\and Biomedical Data Visualization\and Electronic Health Records Visualization}
\vspace{8mm}
  \end{@twocolumnfalse}
]

\begin{tcolorbox}[floatplacement=!b,float,left=2mm,right=2mm,top=1mm,bottom=1mm]
\small
*These two authors contributed equally to this work.\\
This is the authors' preprint version of this paper. License: CC-By Attribution 4.0 International. Please cite the following reference: \\
Haihan Lin, Ryan A. Metcalf, Jack Wilburn, Alexander Lex. 
Sanguine: Visual Analysis for Patient Blood Management.
\textit{Information Visualization}, to appear, 2021.
\end{tcolorbox}

\section{Introduction}
Transfusion is the most commonly performed medical procedure during hospitalizations in the United States~\cite{pfuntner_most_2013}, with 17.2 million blood products transfused in 2015~\cite{ellingson_continued_2017}. Although transfusions are often medically necessary and even life-saving, they have also been identified as one of the most overused procedures in hospitals~\cite{the_joint_commission_and_the_american_medical_association-_convened_physician_consortium_for_performance_improvement_proceedings_2012}. Unnecessary transfusions can lead to adverse reactions and complications, such as volume overload, lung injury, hemolysis, and more~\cite{delaney_transfusion_2016}.

Management and acquisition of blood products is also expensive, creating financial incentives for hospitals to optimize their usage. To better address these problems, many hospitals have hired dedicated experts in patient blood management (PBM), and PBM has become an independent academic field. Through PBM, health care providers aim to reduce unnecessary transfusions and improve patients' outcomes using various methods, such as treating anemia (lack of red blood cells) prior to surgeries, providing guidelines on when transfusions are necessary, and  using cell salvage (``recycling'' blood) during surgery~\cite{gross_patient_2015}. 

Successful PBM involves multiple stakeholders. PBM experts analyze the usage of blood products and give advice or develop guidelines on best practices. Clinicians, in turn, make decisions about when and how much to transfuse, thus they need to be well informed about best practices, their individual performance regarding these best practices, and their performance relative to their peers. 
\begin{figure*}[t]
  \includegraphics[width=\linewidth]{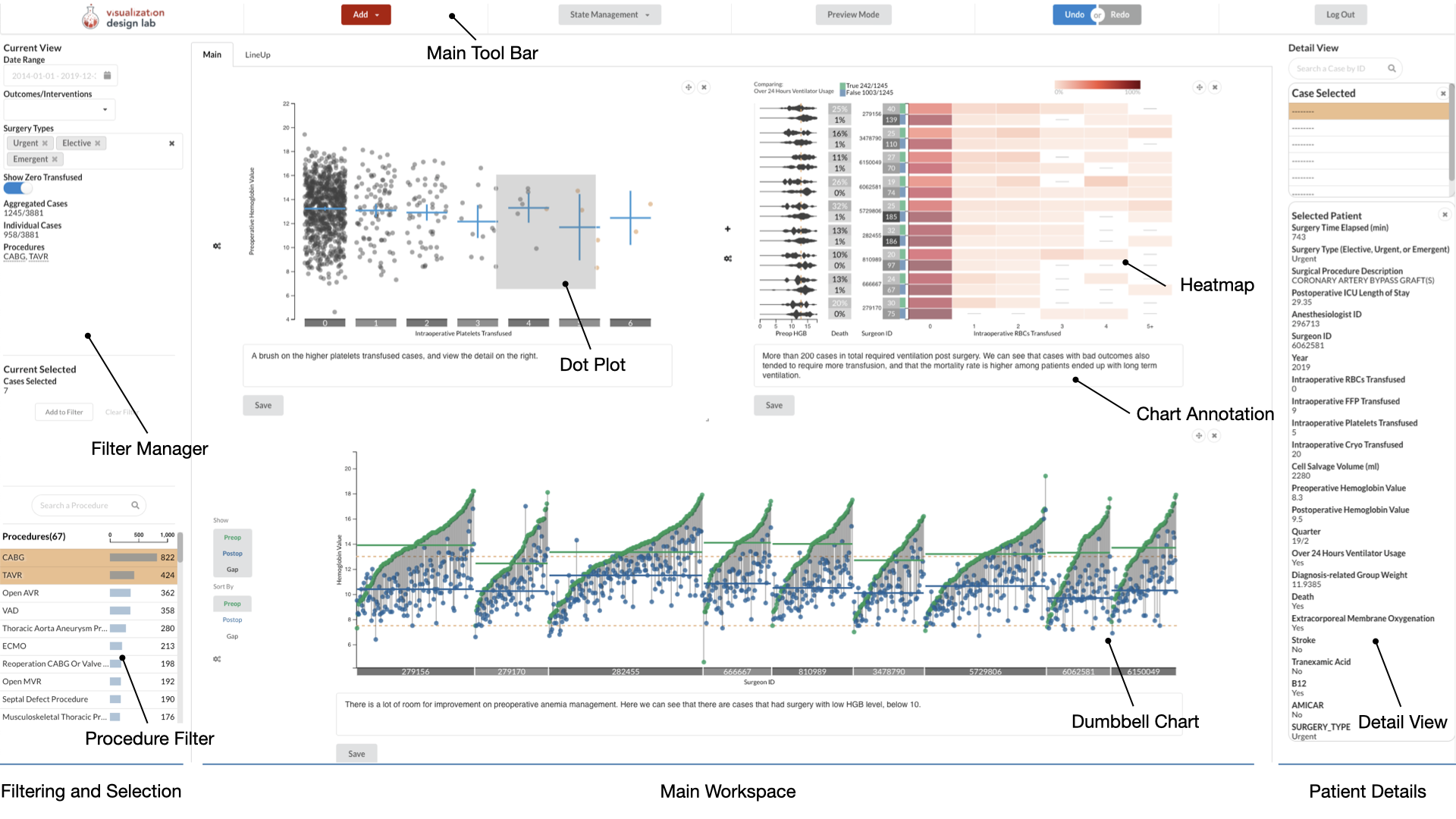}
  \caption{An overview of Sanguine visualizing patient blood management data with multiple views. The left panel is dedicated to managing filters and selections. The workspace in the center contains visualizations that can be flexibly arranged. A heatmap, a dot plot, and a dumbbell chart are shown. On the right, a patient-specific detail view shows attributes of a case.}
\label{fig:sangiune_teaser}
\end{figure*}

These analysis processes, however, are difficult with the current tools available in hospitals. Electronic health records software may provide static charts and reports on blood product usage, but they lack the nuance and sophistication necessary to make a holistic evaluation of the practice of individual clinicians. For example, when one surgeon uses significantly more blood products than their peers, they might rightly claim that they do so because they are treating the most difficult cases with many patient comorbidities. With current tools, analysis is often static or limited to only certain clinical contexts.

To tackle this issue, we created Sanguine, shown in Figure~\ref{fig:sangiune_teaser}, an interactive visual analytics tool for PBM analysis. Sanguine was developed as a design study using an iterative design process involving PBM experts, surgeons, and anesthesiologists at the University of Utah Hospital. Sanguine enables analysts to view outcomes juxtaposed with blood product usage data, review patient blood management practices for different providers, and compare the effect of patient blood management in clinical settings. Sanguine is designed with two types of stakeholders in mind: PBM experts who conduct in-depth analysis, and clinicians (primarily surgeons and anesthesiologists) who receive tailored and annotated interactive reports from the PBM experts. These two groups have radically different types of engagement with Sanguine. The PBM experts independently analyze data and ask a variety of analysis question, creating charts in the process.  The clinicians, in contrast, mostly consume the visualizations as reports that contains feedback on their practice. We designed Sanguine to facilitate communication around patient blood management between the two types of stakeholders. 

Our main contribution is the design and development of Sanguine, an open-source visual analytics tool for the analysis of patient blood management data. We also contribute an analysis of domain goals and a task abstraction. We evaluate the utility of Sanguine in case studies and through interviews with domain experts.

\section{Medical Background}

Patient blood management encourages minimizing blood use through the appropriate use of blood products, ascertained by evidence-based transfusion strategies~\cite{goodnough_patient_2012, shander_bloodless_2012}. However, the decision to transfuse often depends on the clinical experience of the practitioners, and blood product usage can vary significantly among individuals. Analysis of transfusion records~\cite{frank_novel_2013, frank_variability_2012} has shown how different surgeons within the same department vary in their blood use. Current electronic medical record systems do not give practitioners the opportunity to easily review and reflect on multiple cases simultaneously, and the static charts and reports used by clinicians to review transfusions limit them from exploring the collected data in a comprehensive and context-specific fashion.

Blood banks at hospitals regularly report wasted blood products, primarily because of excessive ordering of blood products during preoperative preparations. A blood product issued to the operating room, if unused, must be returned to the blood bank in a timely manner and properly stored; otherwise, it must be discarded. Estimates of wasted blood components range from 200,000 to 1 million in the US annually, with associated costs of 46 million to 230 million US dollars~\cite{hannon_waste_2015}. Many hospitals adopt programs using a maximum surgical blood order schedule (MSBOS). MSBOS contains a list of recommended blood orders for commonly performed surgeries regionally. A ``one size fits all'' solution, however, does not account for much of the variation in patient conditions or risk. A study by Frank et al.~\cite{frank_reducing_2014} has shown that an institution-specific blood order schedule can significantly reduce unnecessary preoperative blood ordering. Metcalf et al.~\cite{metcalf_data-driven_2018} developed an example of a more personalized prediction model to facilitate more accurate and precise ordering, but institutions have not widely adopted this or other novel ordering methods.

\section{Formative Process}

Our main collaborator, Dr. Ryan Metcalf, also a co-author of this paper, is the medical director of the Transfusion Medicine Service at University of Utah Hospital. As a PBM expert, he is studying how transfusion affects patient outcomes and the overall use and effectiveness of patient blood management. Transfusion is common across specialities, including in a nonsurgical context, but we decided to focus on cardiac surgical cases because hemorrhage is highly prevalent in cardiac surgeries. Cardiac surgeries are inherently complex, and numerous PBM modalities exist in this clinical setting.

To get a good understanding of the domain goals, we met with our collaborator on a weekly basis starting in May 2018 about his process for evaluating patient blood management practice. We also involved the data warehouse at the University of Utah Hospital, which aided in curating a dataset, acquired IRB approval for using the anonymized patient data, and secured funding by a university-affiliated non-profit reference laboratory.

Next, to understand the needs of a broader set of stakeholders involved in PBM, in July of 2019, we conducted a creative visualization-opportunities (CVO) workshop~\cite{kerzner_framework_2019}. We recruited participants form Dr. Metcalf's network at the cardiac surgery department. Participants included two cardiac anesthesiologists, one cardiac surgery critical care physician, and one IT manager. Two cardiac surgeons were scheduled to participate but were called away for surgery. An IT manager was included so we could learn more about currently available tools and technology constraints. To accommodate the schedule of the clinicians, the workshop lasted three hours, which is shorter than a typical CVO workshop. After a brief introduction, we used the \textit{Wishful Thinking}, \textit{Visualization Analogies}, and \textit{Barrier Identification and Removal} activities~\cite{kerzner_framework_2019}. The participants identified factors that potentially lead to unnecessary transfusions, and brainstormed on how to improve the current standard of care for better patient outcomes. They agreed on the need for better PBM practices, and were hopeful that a tool showing comprehensive PBM data could achieve this goal.

After the workshop, we did an initial analysis on the result we collected. One key insight that emerged was that clinicians believed that a tailored report to their practice would be much more useful and likely be more widely adopted than a general-purpose tool. This, in turn, let us to design for these two different user groups. To also understand whether the results also match up with cardiac surgeons' workflows, and given that the cardiac surgeon participants had to cancel at the last minute, we decided to present our intermediate results and elicit feedback from a large group of cardiac surgeons during the surgery department's weekly all-hands meeting.  The surgeons stated that a tool that can aid them in their blood management practice could lead to improved outcomes but cautioned that they would not want a system that would obstruct them from ordering blood or slow down their current practice.

The workshop and notes from the all-hands meeting generated rich artefacts, in the form of thematically grouped sticky notes, audio-recordings, photographs, and notes by the facilitators. To analyze, we first created a machine-readable corpus (transcribing sticky notes, etc.) and then grouped the concepts into thematic areas. We then based our domain goals, and subsequently, our design on these results.

\section{Domain Goals}

Based on the CVO workshops, the all-hands meeting, and the regular meetings with our main collaborator, we identified the following high-level domain goals our collaborators have regarding patient blood management:

\begin{itemize}
\item \textbf{Identifying problematic transfusion practices}, and thereby improving outcomes is the primary goal that all our stakeholders share. Meeting this goal requires the analysis of transfusion data in the context of patient records and evidence-based guidelines from the literature. Also, comparisons of various aspects (between clinicians, between time intervals, etc.) emerged as a way to achieve the goal.  
    
\item \textbf{Decision support for when to transfuse} during surgery was brought up as another goal during the workshop. Clinicians would like to have more data and information about when it is appropriate to transfuse while in the operating room.   
    
\end{itemize}

When considering these goals, we quickly realized that real-time decision support would not be possible given the regulatory environment and the data we have available. The US FDA regulates devices (including software) that are used in the diagnosis, mitigation and treatment of diseases. Getting FDA approval for a real-time decision system would have been beyond the scope of this project. Also, the data used for real-time decision making is collected during surgery using various devices and is not stored in real time in a database. The data we can get access to, in contrast, is medical record data. Hence, we decided to focus on the retrospective analysis of transfusion practices.

Based on the collected data and in working with Dr. Metcalf, we identified a typical workflow: He may start by analyzing transfused units by patient preoperative hemoglobin value. A patient should first be treated for anemia if it is present, and ideally have a hemoglobin value above 13g/dL before surgery. If a patient has a hemorrhage during the procedure, cell salvage should be considered as an alternative to transfusion. Our collaborator also analyzes the patient's postoperative hemoglobin value: a hemoglobin value well above the post-transfusion hemoglobin target for transfused cases indicates overuse of red blood cell transfusion. The most important measures are patient outcomes, which can be influenced by transfusion practice. Analyzing transfusions within the context of preconditions, risk factors, details about the surgery, vital signs, and outcomes, is critical for an accurate analysis and can also demonstrate the value of PBM.

Dr. Metcalf described certain barriers to sharing insights effectively with surgeons, which has been only partially successful in the past. Surgeons may not always have the time to analyze data closely. At the same time, they would like for analyses to account for variation in patient characteristics and other specific contexts to meaningfully capture the complexity of a particular case or group of cases. When advising surgeons and anesthesiologists, he would prefer a medium that flexibly and rapidly conveys the key aspects in the appropriate context. Because a well-designed visualization can communicate insights from the data with ease, we decided to design and implement a visual analysis tool for analyzing and reviewing patient blood management practice with outcomes of interest. 

Based on this goal, we curated a list of data items that we would need to address the analysis questions. In cooperation with the data warehouse at the University of Utah Hospital, we compiled a dataset drawing from the electronic health records of over 4000 cardiac surgery patients, spanning the years 2014 to 2020, and covering 111 different cardiac surgery procedures, ranging from routine coronary bypass surgeries to heart transplants. For each patient, we included transfusion records, medication records, and lab results, such as hemoglobin level records and thromboelastogram (ROTEM) results. We break down the high-level goal of identifying problematic transfusion practices into domain goals as follows:

\begin{itemize}

    \item \textbf{G1: Comparing Transfusion Practice.} We identified comparison of transfusion practice as a central aspect of the analysis question. Specifically, our collaborators want to compare transfusion practices among (1) surgeons, (2) anesthesiologists, and (3) years. The former two enable a comparison between individual clinicians, the latter enable comparison over time, to, for example judge the effectiveness of an intervention. However, as was evident from the workshops, crude measures for comparisons, such as total units of blood transfused, do not adequately account for the complexity and heterogeneity of surgical cases. For example, a comparison has to account for the types of procedures (certain procedures rarely require transfusion, while others can result in significant bleeding). A comparison also has to account for other risk factors, such as the general health of the patients. Our interviews and workshop revealed that providing this contextual information is critical for practitioners to trust the visualization. 
    
    \item  \textbf{G2: Analyzing Adherence to Best Practices and Standards.} The fields of cardiac surgery and patient blood management have developed a suite of evidence-based clinical guidelines and best practices for many procedures. For example, postoperative hemoglobin values clearly above a post-transfusion target can indicate over-transfusion, and preoperative hemoglobin values below a threshold can indicate a failure to manage anemia prior to elective (scheduled) surgeries. Similarly the absence or presence of methods, such as recycling the patient's own blood or treating bleeding pharmaceutically are important indicators when analyzing blood management. Both patient blood management experts and clinicians need to analyze whether these standards are met. 
    
    \item  \textbf{G3: Analyzing Individual Patients.} In certain situations, aggregate or summary information does not provide enough information to judge, e.g., why a specific case was an outlier, or why best practices were not followed. To understand the specific situation, clinicians need to access individual patient records.
    
    \item \textbf{G4: Prepare for Surgery using a ``Patients Like Mine'' Approach.} Ordering an appropriate amount of blood is important for both outcomes and efficient use of resources. PBM professionals want to make good estimates for the amount of blood to prepare, taking into account various factors such as the type of surgery, but also key variables about the patient. To do this, they want to explore similar historical cases to see the amount and variability of blood needed, and then order units of blood based on this information.
    
    \item  \textbf{G5: Communication and Sharing.} Finally, since we have identified two types of stakeholders, our visual analysis interface needs to be tailored to both groups: \textbf{PBM professionals}, who want to interactively explore the data, create custom views, and create cohorts based on multiple attributes, whereas \textbf{clinicians} want to see their performance relative to their peers and to analyze whether they follow evidence-based standards. Hence, we envision split roles of a ``power user'' analyzing the data in detail, and providing context through annotations, and a ``consumer'' of these visualizations.
\end{itemize}

These domain specific goal map to a wide variety of visualization tasks. Using Brehmer and Munzner's task typology~\cite{brehmer_multi-level_2013}, G1 to G3 map to the ``discover'' and ``lookup'' tasks on the ``why'' dimension, and the ``encode'', ``select'', ``filter'', and ``aggregate'' tasks on the ``how'' dimension. G1 is a comparison task, for which Gleicher has developed a set of considerations~\cite{gleicher_considerations_2018}. The characteristics of the data subsets to be compared (a handful of different items, such as the records of surgeons, and multiple complex distributions as data values) suggest ``juxtaposition'' as an appropriate design choice for the comparison task.

G4 --- looking for ``patients-like-mine'' is, again using Brehmer and Munzner's typology, of type ``browse'' in the ``search'' category, as the characteristics are know, but the target is not. G5 --- communication and sharing --- is different from the other tasks, as it is mostly about ``producing'' (why) based on ``annotations'' and ``recordings'' (how).

The heterogeneity of the goals and tasks suggests that a flexible, multi-view visualization will be necessary to address all of them.

\section{Related Work}
 
As digital medical records are now ubiquitous, visualizing medical records has been an important area of research for analyzing medical data. We provide an overview here, but for a more detailed survey on visualizing and exploring digital patient records, we refer to Rind et al.~\cite{rind_interactive_2013}. Caban et el.~\cite{caban_visual_2015} describe four major types of visual analytic applications in healthcare: clinicians analyzing patients' records, administrators making data-supported decisions, researchers working on large medical datasets, and patients understanding their own data. Our work falls into the first category. Shneiderman et al.~\cite{shneiderman_improving_2013} analyze the challenges of implementing interactive visualizations for healthcare professionals, which include offering busy clinicians timely information in the right format. West et al.~\cite{west_innovative_2015} encourage creators of visualization systems to also consider the training time required. In Sanguine, we address these points by enabling a workflow where a PBM professional curates an annotated visual report for the clinicians. 

In many clinical contexts, a timeline of symptoms and treatments is informative; hence, multiple works have been focused on visualizing medical records in a timeline format. Lifeline~\cite{plaisant_lifelines_2003}, for example, shows details of individual patient medical histories, representing events and episodes using horizontal lines. TimeLine~\cite{bui_timeline_2007} uses a similar event-based arrangement for personal medical records, but is abundant in image and file resources for users to explore. Faiola et al.'s work~\cite{faiola_medical_2014} is specifically designed for ICU clinicians, juxtaposing event timelines over vital signs. DecisionFlow~\cite{gotz_decisionflow_2014} visualizes a group of patients with similar characteristics in temporal event sequences. Outflow~\cite{wongsuphasawat_exploring_2012} also provides an overview of event sequences extracted from groups of patient records. All these works focus on visualizing an individual patient or a group of patients on a timeline, and emphasize specific events happening in the sequence. While event sequences can play a role in transfusions, Sanguine takes a provider and practice-focused approach instead. 

Patient cohorts analysis is another mainstay of visualizing medical records~\cite{shneiderman_improving_2013}. Cohort definition and creation is a common theme in these works. COQUITO~\cite{krause_supporting_2016} uses a treemap and visual temporal queries to help users generate cohorts. Composer~\cite{rogers_composervisual_2019} adopts an attribute filter method for cohort creation, and compares patient cohort developments and changes over time and to other cohorts. Bernard et al.~\cite{bernard_visual-interactive_2015} also follow the attribute-oriented cohort creation, but visualize all patient records of the selected cohort, instead of comparing several cohorts through an aggregated visualization design. Phenostacks~\cite{glueck_phenostacks_2017} uses a set visualization to compare various phenotypes over multiple cohorts simultaneously and a sunburst graph for subset and global pattern discovery. Bernard et al.~\cite{bernard_using_2018} use static dashboards and compact data encodings to visualize groups of patients; their approach lacks the flexibility for users to curate their own charts. Visualization studies of patient cohorts analysis typically focus on cohorts that are defined by patient characteristics, such as shared demographics. Sanguine, instead, focuses on cohorts that relate to the practice of the providers: cohorts are either created temporally, to make year-to-year comparisons, or by the provider, to compare the performance of surgeons and anesthesiologists. 

Medical record visualizations can help users navigate large datasets. Lee et al.~\cite{lee_web-based_2016} developed a web-based visualization tool to help users querying large datasets with comparative visualizations; however, the tool has limited ability for customization. 

Another research topic on visualizing health record is bridging the communication gap between providers and patients~\cite{rajwan_medical_2010}. As an example, PROACT~\cite{hakone_proact_2017} helps patients understand the complex risk of cancer and the treatment plans available to them through visualizations. We are not aware of existing techniques explicitly designed for provider-to-provider communication, which is a key aspect of Sanguine.

Finally, various projects have applied statistical analysis to transfusion data. Gálvez et al.~\cite{galvez_visual_2014} developed a tool for decision-making support on blood product ordering at a pediatric hospital, which gives patient-specific blood order advice. This tool allows users to filter the data based on age groups, procedures, and conditions. After filtering the data, the tool visualizes all the transfusion data and other attributes of interest in previous procedures using bar charts. This approach offers a concrete solution to a patient-specific blood order schedule, and shows how visualization can be effective for visualizing transfusion in a clinical setting. Although the tool shows all blood usage of cases from filtered results, Sanguine offers more personalized blood ordering preparation for providers, where they can view their history of transfusions on a particular procedure with specific conditions. 

\section{Visualization Design}

\begin{figure}[t]
    \centering
    \includegraphics[width=0.9\linewidth]{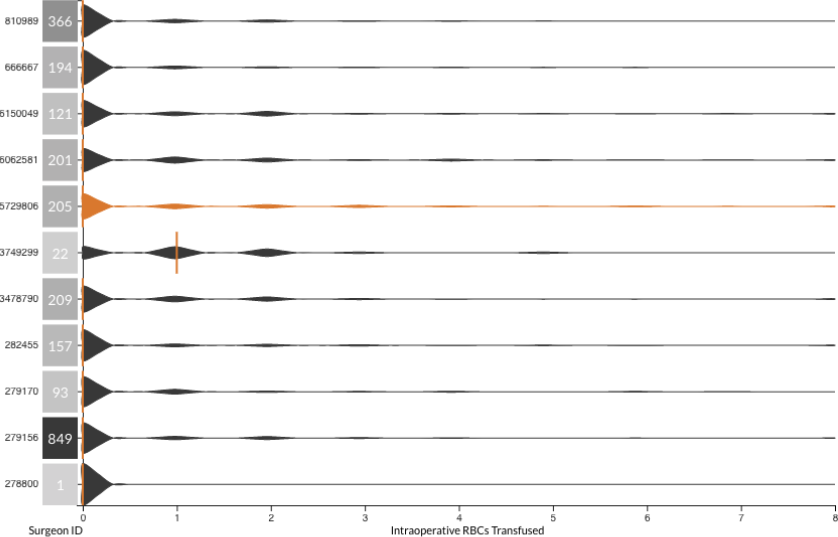}
    \caption{Our initial design used violin plots to show distributions. We abandoned this design due to the skewed and discrete distribution of the data.}
    \label{fig:violinplot_design}
\end{figure}

\begin{figure*}[t]
 \centering 
 \includegraphics[width = \textwidth]{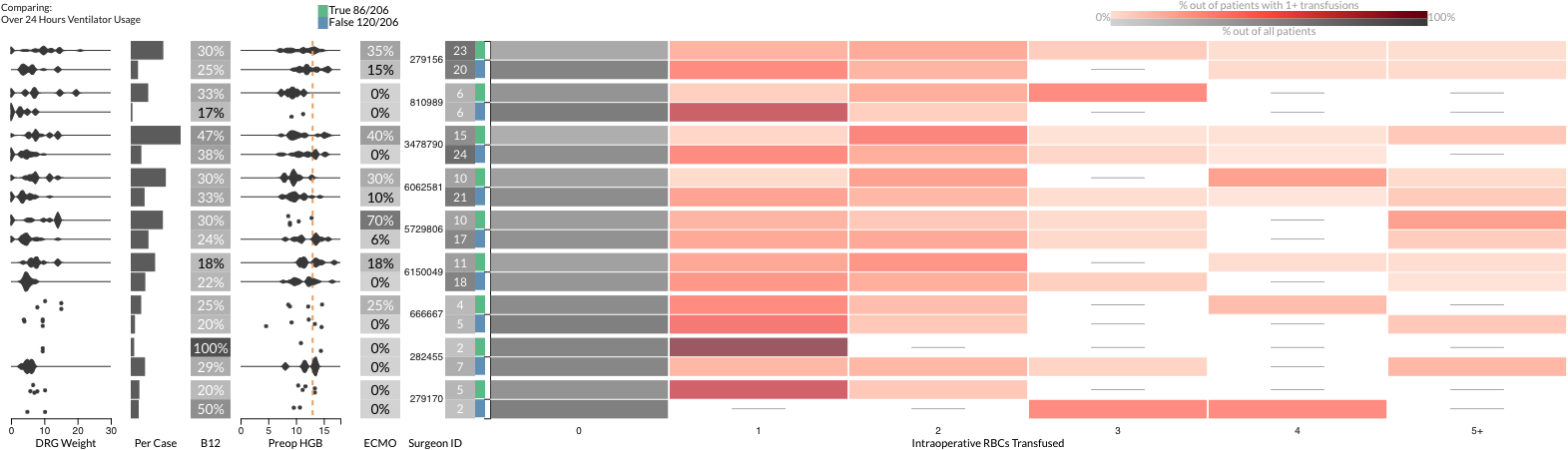}
 \caption{A heatmap showing transfusion data (red cells) for surgeons performing bypass procedures (CABG), subdivided by the need for long-term ventilation. Cases shown in the rows with a green indicator required ventilation for more than 24 hours, whereas cases in blue rows were removed from ventilation before that, or did not require a ventilator. For context, five attributes are visualized on the left: distribution of DRG weights (risk scores), per case transfusion amount, B12 medicine usage, preoperative hemoglobin distributions, and ECMO usage rates. The red color map for red blood cell transfusion is set to exclude cases without transfusions, and the gray color map shows the percentage of zero transfusion in the total cases of each row.}
 \label{fig:outcomes}
\end{figure*}

The design of Sanguine reflects the goals described in \textbf{Domain Goals} section, which, in turn, are rooted in the interviews and creativity workshop we conducted. The subsequent development of Sanguine was done using an iterative, user-centered process based on the feedback from weekly meetings between the visualization development team and the PBM expert. Because we are designing this visualization for two user types, Sanguine needs to be powerful enough for PBM experts to discover relevant results, and be straightforward enough so that surgeons can leverage the visualization with limited time.

Sanguine consists of three main components, shown in Figure~\ref{fig:sangiune_teaser}: a filtering and selection view, the main workspace where visualizations can be flexibly arranged, and a patient-specific detail view. A video demonstrating Sanguine is available at \url{https://youtu.be/DhTNyvCJgtM}. The source code is available at \url{https://github.com/visdesignlab/Sanguine}. Due to the sensitive nature of the data, we cannot provide a live-demo.

\subsection{Data}

A large amount of data is recorded in patient records, but not all of the data is typically necessary to answer a particular research question. Hence, we designed Sanguine such that a PBM expert can flexibly compose visualizations based on the data relevant to the task. The primary data that PBM experts are interested in are the number of units transfused, cell salvage (an alternative to transfusions, measured in milliliters), and laboratory measurements of hemoglobin level, which are the main indicators for anemia management and transfusion appropriateness.

Depending on the research question, other data can provide relevant context: patient outcomes (death, need for long-term ventilation, etc.), PBM-related drug administration, and information about whether a surgery was planned or was an emergency are examples. These attributes are extracted directly from electronic health records and anesthesia flowsheets.

\subsection{Comparing and Contextualizing}

The most important functionality of Sanguine is to visualize the transfusion data mentioned above, while addressing the comparison-related goals --- comparing between practitioners and time intervals (\textbf{G1}), as well as dividing and comparing using additional variables, such as outcomes. Breaking these goals into abstract tasks, we need to enable the comparison of distributions of values. Following Gleicher's guidance~\cite{gleicher_considerations_2018}, we chose a juxtaposition strategy. 

In the early development stage, we used bar charts to show the total number of transfusions, but quickly decided to abandon bar charts because they did not account for the number of surgeries performed, or for the number of units transfused per surgery. We then evaluated violin plots (Figure~\ref{fig:violinplot_design}), since violin plots are generally well suited to visualize distributions. However, as is evident from Figure~\ref{fig:violinplot_design}, the discrete nature of the data made violin plots a poor choice for visualizing these values.

Hence, we decided to use heatmaps as the main chart type for comparison visualization, as shown in Figure~\ref{fig:sangiune_teaser}. Addressing the practice-focused comparison (\textbf{G1}), the heatmap shows transfusion data aggregated by surgeons, anesthesiologists, or years. To account for the difference in total case count between these facets, we use relative scales showing the percentage of cases that had 0--4 units of red blood cells, or 5 or more transfusions. The threshold varies depending on the blood component. The decision to aggregate values above a certain threshold into one bin was driven by the effect that rare outliers had on the overall visualization. For cell salvage data, which are measured in continuous values, we used bins with a dedicated bin for no usage. 
    
As is evident from the heat map in Figure~\ref{fig:sangiune_teaser}, a large percentage of cases received no transfusion. This aspect of the data is important to include in the visualization, but the heavily skewed distribution makes it hard to notice any difference in the nonzero transfusion cases. To address this problem, we provide a toggle to remove zero transfusion from the color-scale in the visualization, i.e., the maximum value that maps to the darkest red color is taken from the counts of cases that have received at least one transfusion. This makes differences in transfusion practice much more apparent, as is evident in Figure~\ref{fig:outcomes}, which is important as analyzing cases that required transfusion is a key task. The heatmap still shows the data on zero transfusion, but uses a separate (gray) color scale encoding the percentage of zero transfused cases out of all cases of the row. 

The resulting heatmap enables PBM experts and clinicians to compare their relative practice of transfusion (\textbf{G1}). Although Sanguine can be used to compare transfusions across all procedures, filtering by procedures is essential, because diverse procedures such as heart transplants and bypass surgery differ in the typical need for transfusions. However, even when narrowing cases down to specific procedures, there can be systematic differences in how complicated cases are, and hence what outcomes can be expected and how much will be transfused.

To provide this context and show transfusion-relevant outcomes, Sanguine can visualize additional attributes to display along with the heatmap, such as lab values and patient outcomes (see Figure~\ref{fig:outcomes}). The visual encoding used depends on the attribute types. For numerical/distribution values (such as hemoglobin values), we use a violin plot that morphs into a dot plot when there are few observations; for individual numerical values, we use bar charts (e.g., average transfusions per case) or labeled heatmap cells (e.g., mortality rates). 

\begin{figure*}[h]
\captionsetup[subfigure]{justification=centering}
\centering
    \begin{subfigure}[b]{0.36\textwidth}
    \centering
    \includegraphics[width=\textwidth]{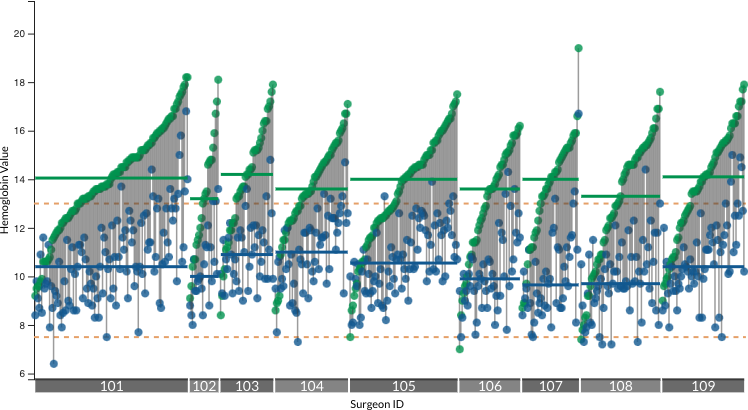}
        \caption{Dumbbell Chart}
    \label{fig:dumbbell_before_filter}
\end{subfigure}
\hfill
    \begin{subfigure}[b]{0.3\textwidth}
    \centering
    \includegraphics[width=\textwidth]{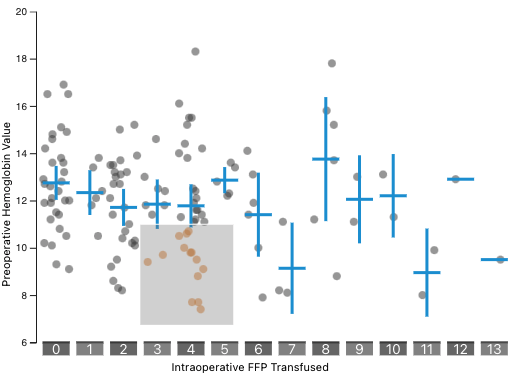}
    \caption{Dot Plot}
    \label{fig:dotplot}
\end{subfigure}
\hfill
    \begin{subfigure}[b]{0.32\textwidth}
    \centering
    \includegraphics[width=\textwidth]{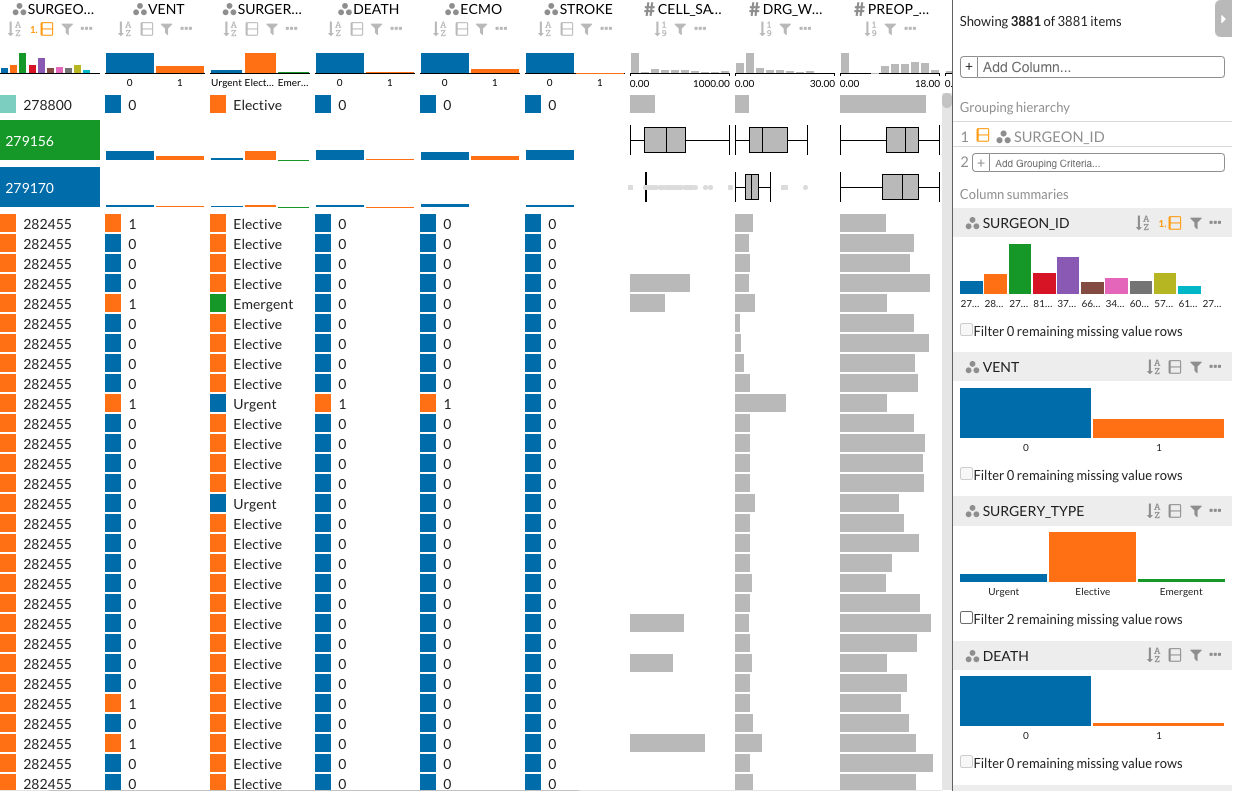}
    \caption{LineUp}
    \label{fig:lineup}
\end{subfigure}
\caption{Visualizing bypass (CABG) procedure data: (a) A dumbbell chart that shows pre- and postoperative hemoglobin values of elective surgery cases, grouped by surgeons and sorted by preoperative hemoglobin levels. Horizontal lines in each band represent the medians. The stippled lines are for references to preoperative hemoglobin level (13g/dL) and transfusion-trigger hemoglobin level (7.5g/dL). (b) A series of dot plot shows the postoperative hemoglobin value grouped by units of frozen plasma transfused intraoperatively. The horizontal line in each band represents the mean, and the vertical line is the 95\% confidence interval. (c) LineUp, a tabular data visualization technique, is available in Sanguine as a separate tab view. The filtering system in LineUp is connected to the main view.}
\label{fig:dot_and_dumbbell}
\end{figure*}

Finally, to enable outcome and intervention-focused comparisons, the heatmap in Sanguine can be divided by binary outcome variables such as mortality, or by a time-range. The heatmap shown in Figure~\ref{fig:outcomes} is divided by the need for long-term ventilation, as indicated by the green and blue fields, respectively. Every row is divided, showing the transfusion data and contextual information in separate rows. In Figure~\ref{fig:outcomes}, for example, we can see that long-term ventilation appears to correlate with higher transfusion rates and higher risk scores. 

\subsection{Standards and Best Practices}

Hemoglobin laboratory values are a key indicator in patient blood management. Clinical trials have led to the development of evidence-based guidelines that use hemoglobin value as red blood cell transfusion thresholds and corresponding expectation for post-transfusion hemoglobin targets can thus be inferred. Hence, analyzing transfusion practice in this context is essential (\textbf{G2}). A low preoperative hemoglobin level indicates improper anemia management before surgery, whereas high postoperative hemoglobin levels imply excessive transfusion during surgery. To facilitate the analysis, Sanguine provides a dumbbell chart dedicated to hemoglobin level evaluation (see Figure~\ref{fig:dumbbell_before_filter}). We encode the recommended levels for pre- and postoperative hemoglobin values in the chart, and they are easy to change in Sanguine through a configuration file. An individual dumbbell visualizes the preoperative (green) and the postoperative (blue) hemoglobin values of a single case as dots. A link connecting the dots shows the gap between these values. Cases and their corresponding dumbbells can be divided by all relevant attributes; Figure~\ref{fig:dumbbell_before_filter} shows dividing by surgeons. Within each division, solid horizontal lines show the medians for pre- and postoperative hemoglobin level respectively, and stippled lines provide clinically recommended values for minimal preoperative hemoglobin level and transfusion-trigger hemoglobin level i.e., the level at which clinicians should initiate transfusion during surgery. The dumbbells within each division can be sorted based on preoperative value, postoperative value, or the gap between the two. 

\subsection{Filters and Brushes}

In addition to dynamic comparisons, filters and brushes are essential tools to create the visualization relevant to answer specific analysis questions. In addition to the procedure filter view (Figure~\ref{fig:sangiune_teaser}), Sanguine provides views specifically designed for filtering based on data values. The dot plot (Figure \ref{fig:dotplot}) can visualize correlations and support rectangle brushes, which can be converted into filters in the filter manager view (Figure~\ref{fig:sangiune_teaser}). The dot plot also visualizes mean values and confidence intervals.   

We also integrate a LineUp~\cite{gratzl_lineup_2013, furmanova_taggle_2020} view (Figure~\ref{fig:lineup}) to visualize all attributes of individual cases (\textbf{G3}) and complement Sanguine's filter system with the one that comes with LineUp. LineUp visualizes data in a tabular layout, and applies different techniques to columns based on the attribute types.

\subsection{Case View}    
    
An alternative way to view information about individual cases (\textbf{G3}) is the detail view, shown in Figure~\ref{fig:sangiune_teaser} on the right. All selected cases are shown in a list; the attributes for one case are shown, including transfusion record, surgery description, relevant medicines administered, and outcomes. Using the detail view, analysts can study these cases closely, and if they identify something of interest, they can add that attribute to another visualization in Sanguine. Ultimately, practitioners can also cross-reference cases with the medical record system, which contains even more data on the patients.

\subsection{Communication and Sharing}

Sanguine enables communication and sharing (\textbf{G5}) in two complementary ways: first, communication is facilitated through annotations: each visualization is accompanied with a text field, which can be used for notes or to record insights and conclusions. 

Second, sharing of findings is enabled through provenance tracking. Each change to the visualization, including visualization configurations, filter settings, and annotations, is saved as a state, and recorded using the Trrack provenance tracking library~\cite{cutler_trrack_2020}. Based on Trrack's functionality, Sanguine provides undo/redo, saving and loading the state of a workspace to/from a server, and sharing the state via a URL. URL based sharing is convenient for distributing findings made by the PBM expert to clinicians, who can then review the visualizations and adjust their practice, if appropriate. As these ``visualization consumers'' typically do not want to leverage the full complexity of Sanguine, we introduced a ``View Mode'', which removes editing functionality and simplifies the interface. 

\subsection{Implementation and Deployment}

Sanguine is developed in Typescript using the React framework. We use Semantic UI and D3 for the graphical interface and the MobX and Trrack library~\cite{cutler_trrack_2020} for state control. The frontend is supported by a Django server. The server interfaces with an SQL database housed in the data warehouse of the University of Utah Hospital. It is deployed in a protected environment suitable for sensitive medical data, and uses encryption and two-factor authentication to ensure security.

\section{Evaluation}

We evaluate Sanguine through case studies with Dr. Metcalf, the PBM expert, and through feedback from surgeons, thereby covering our two user groups of analysts who interact with the tool in depth and clinicians who consume curated reports. 
As our collaborators were closely involved in the development of Sanguine, experimental demand characteristics, which often arise in close collaborations~\cite{brown_into_2011}, can bias responses, hence, we refrain from reporting subjective assessments and instead report factually on analysis scenarios and make comparisons of capabilities of a workflow with and without Sanguine.

\subsection{Case Study}

To demonstrate the utility of Sanguine, we present case studies of two scenarios where a PBM expert, Dr. Metcalf, uses Sanguine (1) for analysis of transfusion practices, and (2) as a ``patients like mine'' decision support tool when ordering blood components during preparation for surgery. The main research question for these case studies is if Sanguine can provide insights into transfusion data and can be used to answer the domain goals. Our collaborator had the opportunity to use the tool in practice for several months after we completed a prototype.

\subsubsection{PBM Review and Analysis}

\begin{figure}[h]
 \centering
  \vspace{-4mm}
 \includegraphics[width=\linewidth]{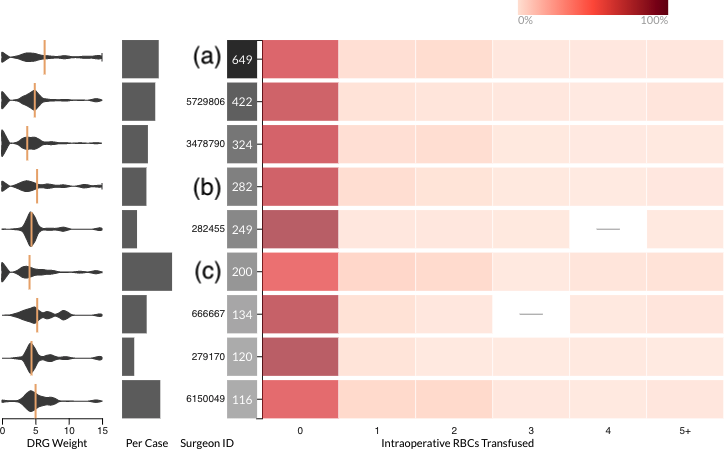}
 \caption{Heatmap of units of red blood cells transfused for the CABG, TAVR, Open AVR, VAD, and thoracic aortic aneurysm procedure. DRG weights and per case units transfused are added for peer group benchmarking. Surgeon (a) and (b) have higher medium value for DRG values,  6.39 and 5.26 respectively, but their per case units transfused are 0.67 and 0.45, whereas cases treated by surgeon (c) have relatively low average DRG weight (4.13), but a high transfusion rate at 0.93 units per case.}
 \label{fig:peer_group_benchmark}
 \vspace{-2mm}
\end{figure}

\paragraph{Peer Group Benchmarking.}
Context matters when comparing the transfusion practices of surgeons and anesthesiologists to their peers (\textbf{G1}, comparing transfusion practice). For example, if one surgeon transfuses more blood on average than others in the peer group, the surgeon may rightly complain that the comparison is not fair if they tend to see sicker or more complex patients. Previous work~\cite{metcalf_association_2019} has shown that diagnosis related group (DRG) billing code weights are associated with transfusion volumes in surgical populations. In other words, more complex patients will have a higher DRG weight and would be expected to receive more blood on average.

To analyze differences in transfusion practice, the PBM expert created a heat map to visualize blood utilization of the different surgeons, shown in Figure~\ref{fig:peer_group_benchmark}. He filtered to include only the most common complex cardiac surgery procedures. Blood use varies substantially, so to see whether the differences was due to the complexity of the cases and allow for fairer comparisons, he chose to visualize the distribution of DRG weights next to the heat map and the bar chart of average transfusion per case. By adding the DRG weight distribution, it is easy to see which surgeons represent a reasonable peer group comparison and which surgeons do not. In the example shown in Figure~\ref{fig:peer_group_benchmark}, the two surgeons (indicated as (a) and (b)) seeing the most complex cases did not have the highest blood utilization, while one surgeon (c) mostly saw cases with low DRG weight but still transfused a lot. The expert commented that this identifies surgeons performing above and below of the context-appropriate peer group, which in turn can be used as a starting point for a discussion and intervention to improve practice.

\paragraph{Anemia Management.} Preoperative anemia management (\textbf{G2}, best practices) is an essential part of patient blood management. Dr. Metcalf started his analysis by studying whether patients who had anemia were treated for it before surgery, as anemia can require transfusions that would otherwise not be necessary. A patient with a hemoglobin level at 10g/dL or below is considered to be anemic. To view the overall hemoglobin trend, our collaborator first created a dumbbell chart comparing surgeons. Since preoperative anemia management can be done only for elective surgeries, he then used the surgery urgency filter in the filter manager to remove emergent and urgent cases. He also selected a common procedure coronary artery bypass grafting (CABG) in the filter selection, and analyzed a subset of surgeons who have performed CABG. By comparing preoperative hemoglobin values to the stippled line for 13g/dL as the clinically recommended preoperative hemoglobin values, as shown in Figure~\ref{fig:dumbbell_before_filter}, he noticed that several cases should have been treated for anemia before surgery, but were not. He also noticed that there appeared to be some variation in patient preoperative hemoglobin levels between surgeons.

\begin{figure}[h]
\centering
 \includegraphics[width=\linewidth]{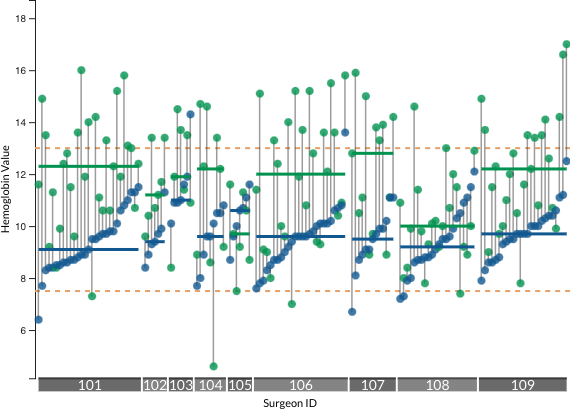}
 \caption{Transfusion Appropriateness: A dumbbell chart comparing surgeons, sorted by postoperative hemoglobin levels. Cases shown are with one or more units of red blood cell transfusions. The high postoperative hemoglobin values (blue line above the lower stippled line) indicate that over-transfusion is widespread.}
 \label{fig:dumbbell_after_filter}
\end{figure}

\paragraph{Transfusion Appropriateness.} Using another dumbbell chart, Dr. Metcalf also reviewed transfusion appropriateness (\textbf{G2}, best practices). When a patient receives a red blood cell transfusion, the provider's target postoperative hemoglobin value is between 7g/dL and 9g/dL. To view only patients who received red blood cell transfusions, our collaborator added a dot plot of postoperative hemoglobin and units of red blood cells transfused intraoperatively. Appropriate transfusion is a practice applicable to all cases, elective or urgent; hence, he removed the elective case filter, and sorted the dumbbell chart to postoperative hemoglobin levels for easier comparison of the postoperative values. Using the brush feature of the dot plot, our collaborator could filter out patients who did not receive red blood cell transfusions during the surgery, resulting in the dumbbell chart shown in Figure~\ref{fig:dumbbell_after_filter}. After applying the filter, our collaborator observed a significant amount of cases for which postoperative hemoglobin values were much higher than the target value, indicating over-transfusion in these cases. He then used this information to advise surgeons and anesthesiologists for better hemoglobin level targeting when performing transfusions.

\paragraph{Cell Salvage.} Another key point of patient blood management is the use of cell salvage. Cell salvage is recycling a patient's blood when they bleed during the surgery, an alternative to transfusions. Ideally, all surgeons should use cell salvage when performing transfusions (\textbf{G2}, best practices). Dr. Metcalf started with a heatmap of cell salvage, which he aggregated by surgeons. The heatmap shows that most surgeons used cell salvage, but the chart does not indicate how surgeons were using cell salvage when they were also transfusing.

\begin{figure}[h]
 \includegraphics[width=\linewidth]{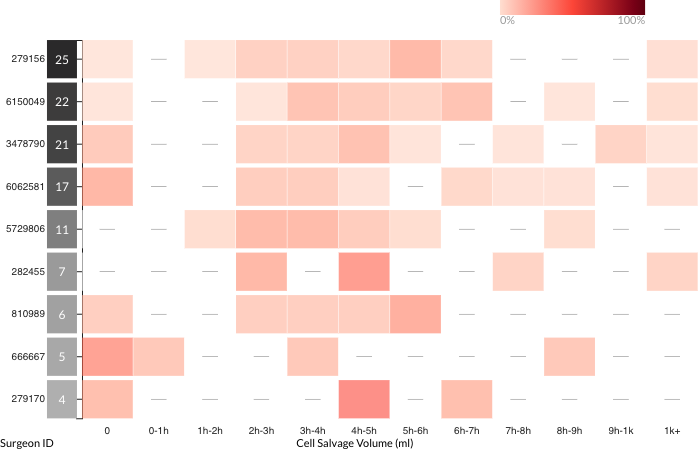}
 \caption{Cell Salvage Usage: A heatmap of cell salvage usage, for cases with one or more units of red blood cells transfused. Several surgeons do not or rarely use cell salvage when transfusing, which violates best practices }
 \label{fig:cell_salvage_with_filter}
\end{figure}

After removing all cases without transfusions, our collaborator observed that even though cell salvage was used in most cases, there was still room for improvement, as shown in Figure~\ref{fig:cell_salvage_with_filter}. Most providers were using cell salvage properly; however, a few providers were performing transfusions without using cell salvage for over 20\% of their cases. This information is valuable to PBM experts, who can now identify the providers and inform them about properly using cell salvage. 

\begin{figure}[h]
\centering
 \includegraphics[width=\linewidth]{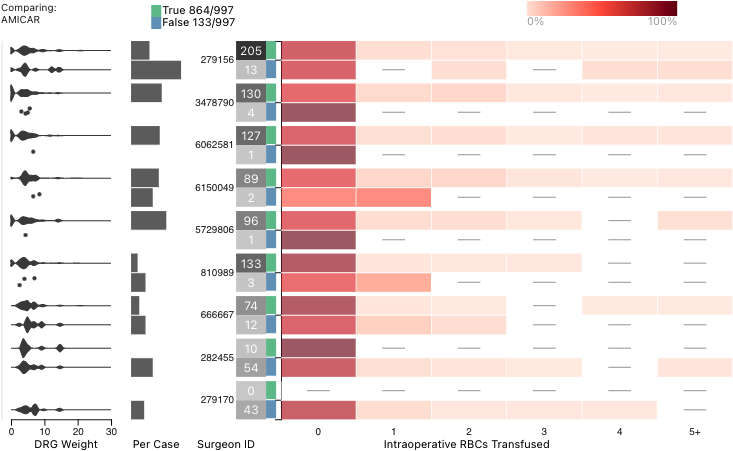}
 \caption{AMICAR Usage: A heatmap of red blood cell transfusion, divided by the use of aminocaproic acid, a drug that can reduce bleeding. The violin plot shows DRG weights (risk scores), and the bar chart shows the units transfused per case. Cases using aminocaprioc acid used less RBC units.}
 \label{fig:amicar_comparison}
\end{figure}

\paragraph{Drug Treatment.} Finally, Dr. Metcalf analyzed usage of antifibrinolytic agents, in this case, aminocaproic acid. Antifibrinolytic agents help to reduce bleeding in surgeries, and overall reduce the use of transfusion (\textbf{G2}, best practices). To visualize the use of aminocaproic acid, he added a comparison heatmap of red blood cells transfused, aggregated by surgeon, divided by aminocaproic acid usage, shown in Figure~\ref{fig:amicar_comparison}. The chart gave him an idea of how frequently providers were administering aminocaproic acid, and how effective it was at reducing transfusions. By analyzing the heatmap and comparing the case counts, he concluded that the majority of cases did appropriately receive aminocaproic acid, which is indicated for patient blood management in cardiac surgeries. 

Our collaborator identified a trend of fewer transfusions associated with the use of aminocaproic acid across individual surgeons To investigate the use of aminocaproic acid in more context, he added plots of the distribution of the DRG weights (risk scores). He noticed that despite similar DRG weights, cases using aminocaproic acid used less blood. This is consistent with what is known about the benefit of this drug in cardiac surgery (reduced need for transfusions) and further drives home the importance of this patient blood management modality.

Dr. Metcalf remarked that none of these analyses were possible before, at least not with the flexibility that Sanguine enables.

\subsubsection{``Patients Like Mine'' Decision Support}
\begin{figure}[h]
 \centering
 \includegraphics[width=\linewidth]{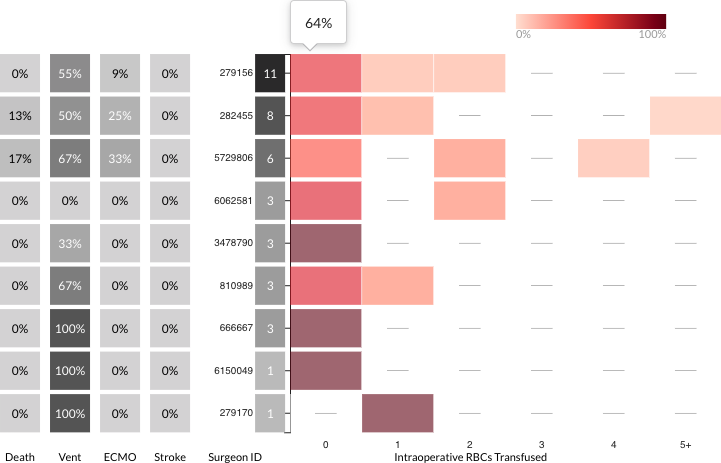}
 \caption{Heatmap of units of red blood cells transfused for the open MVR procedure and cases with preoperative hemoglobin levels of 10-12g/dL. By narrowing the records down to procedures and parameters specific to the case, surgeons can plan how many units of blood to order.}
 \label{fig:Open_MVR_Hemo_Filtered_Heatmap}
\end{figure}

\hfill\\
Dr. Metcalf explored using Sanguine as a decision support tool for blood component ordering when preparing for surgeries (\textbf{G4}). For example, a surgeon preparing for an open mitral valve replacement (MVR) can view the transfusion data for prior open MVR cases, using Sanguine's heatmap shown in Figure~\ref{fig:Open_MVR_Hemo_Filtered_Heatmap}. To get a more specific picture, they can apply a filter using the dot plot based on their patient's preoperative hemoglobin value, and even filter based on the anesthesiologist with whom they will work in the surgery. The surgeon can view their own historic transfusion record for open MVR and order blood components based on their personal records. For example, a surgeon who performed 11 open MVR cases (first row in Figure~\ref{fig:Open_MVR_Hemo_Filtered_Heatmap}) did not transfuse in 64\% of cases, and used only 1 unit in 15\% of all their cases. Hence, they can conclude that they should order one unit of red blood cells before surgery. Surgeons can also view the outcomes of cases from historic data, and know what they can expect for their upcoming case. For example, for the same surgeon in the first row, they can see that 55\% of the cases required long-term ventilation after the surgery. 

\subsection{Feedback from Clinicians}

Dr. Metcalf, the PBM expert, presented his findings, described in the previous section, to a cardiac surgeon and several anesthesiologists in virtual meetings, using a screen-share of Sanguine in ``View Mode'' (G5, communication). As we designed Sanguine to be used by clinicians as visualization consumers, we considered the modality of the PBM expert presenting via screen-share and advising on blood product usage representative of a real-world use case.

 After the presentation, the surgeon said, \textit{``this is what I have been looking into and could not find an answer anywhere. Besides just blood transfusion, there is a lot of other data in there too.''} The anesthesiologists discussed opportunities Sanguine could bring for implementing interventions:  \textit{[Sanguine] can guide our potential interventions...this is great to look at combinations of data...and for evaluating practices. I would use this tool every day.}

We asked them to compare Sanguine with their current practice for analyzing PBM data, which consists mostly of reviewing the semi-annual report provided by the Society of Thoracic Surgeons. The surgeon commented on the difference: \textit{the STS reports we currently receive are about six months delayed, and you can update [Sanguine] much faster... This can be used for meetings to see where we can improve.} 

Speaking to a weakness, the surgeon raised a concern regarding the integrity of some of the data. First, they questioned the validity of the cell salvage values, because sometimes the usage is not properly captured in the anesthesia flowsheets, from which Sanguine derives the data. They also were surprised by and doubted the accuracy of the initial database results showing minimal usage of aminocaproic acid, which they expected everyone to use on every case for cardiac surgery. This allowed us to go back to the database and identify an error in how the aminocaproic acid data was extracted and populated. This initial error was present in the data that Sanguine is built on, but was subsequently resolved after our investigation that spanned several month with the data warehouse. The observation that the data that is present in a database may not be a perfect reflection of reality, and that domain experts often have deep knowledge about such issues \cite{mccurdy_framework_2019} spawned a line of future work our lab is currently pursuing: How can such knowledge be externalized and made explicit in the visualization so that it can be shared and reviewed. 

\section{Discussion and Limitations}

Overall, Dr. Metcalf remarked that using Sanguine is a much more personalized and precise way for surgeons to prepare, compared to the guidelines currently in place. He commented that in this way, Sanguine can help reduce waste, while at the same time also reducing the need for emergency blood releases during the surgery. The COVID-19 pandemic has led to unprecedented blood shortages at hospitals. Each morning, our collaborator has been using Sanguine's ``patient like mine'' support to predict blood utilization for the most complex surgeries for the day. This, in turn, has enhanced preparation for individuals surgeries and to maintain the hospital blood bank inventory to meet critical patient needs.

\paragraph{Comparison to Prior Practice} Prior to the implementation of Sanguine, the most commonly used information about clinical practice quality at our institution are reports from the Society of Thoracic Surgeons (STS). 
The STS database includes quality reporting data and has many participating sites across the United States. However, these data are static and not focused on patient blood management (PBM), and there is typically a time lag of several months to receive reports. Other institutions have published their standard PBM performance visualizations in the clinical literature~\cite{frank_novel_2013}. While these are useful for evaluating broad trends, they lack detailed context and are not flexible or interactive like Sanguine.
In contrast, Sanguine’s dynamic and interactive nature allows the user to interact with the data and create highly contextualized visualizations. The filtering and context setting adds meaning because the clinical circumstances are important for understanding what an outcome means or why it may have occurred. 
However, a benefit of the STS report over Sanguine is that they can also be used to make comparisons between different clinics. We hence believe that Sanguine and STS reports complement each other well: the former is useful for internal analysis and benchmarking, the latter can be used to compare the overall performance of a site nationally.

\paragraph{Evaluation.} We demonstrate the utility of Sanguine through case studies and through feedback from clinicians and PBM experts. Case studies with real users, tasks, and datasets are the most common evaluation approach in design study research~\cite{sedlmair_design_2012}, and demonstrate rich, practical and impactful insights on the domain. Even though case studies tend to not significantly contribute to an understanding of the research contribution, they are considered necessary to ensure validity and trust~\cite{rogers_insights_2021}. Nevertheless, our validation is limited to our primary collaborator and a small group of clinicians. Alternative evaluation strategies, such as a comparative, quantitative evaluation, are difficult to achieve with specialized tools due to the lack of alternative systems that address the same task, and the small group of expert users that could meaningfully participate. We consider long-term observations on usage and adoption as the gold standard for evaluating design study tools. Sanguine has now been used in regular practice for about a year by Dr. Metcalf. However, adoption by outside, unaffiliated parties takes time, as striking agreements, getting IRB approval, and compiling the relevant databases in a different hospital is a complicated process. While we are currently discussing adoption in two unaffiliated medical centers, reporting on adoption in these centers is beyond the scope of this paper. 

\paragraph{Data Integrity.} As described in previous section, the integrity of some of the data was questioned when we presented Sanguine to clinicians, which led to the identification of a problem in how the database was populated. A tool like Sanguine can make the clinical team aware of deficiencies in the data collection process. However, this problem indicated that it will be important for Sanguine to provide a better annotation system or even more innovative techniques to record these disagreements with the data, enabling PBM experts and clinicians to pinpoint and communicate these problems. 

\paragraph{Data Integration.} The dataset we used to develop Sanguine is static and deidentified; however, the deidentification procedures are not certified and, hence, we cannot show certain data fields in this paper. Since we query the data directly from the data warehouse of the University of Utah Hospital through SQL queries, and the data types were not altered during deidentification, we expect a smooth transition from deidentified dataset to the original dataset for clinical use.

\section{Conclusion and Future Work}

In this paper we introduced Sanguine, a tool for analyzing patient blood management data. Sanguine is currently deployed and our collaborator has adopted it in his regular workflow. Going forward, we plan to evaluate the long-term use of Sanguine and also study the influence it has on the decision-making process with independent experts at different institutions. We are currently in negotiations to deploy Sanguine in two different hospital systems. However, due to the necessary customizations, as data are stored differently in each hospital, we expect that such a deployment and subsequent evaluation may take up to a year. Nevertheless, we consider the willingness of these outside hospital systems to invest in deploying Sanguine as an indication for its usefulness beyond our own institution. 

While we have taken first steps towards a dedicated, simplified interface for clinicians with the ``View Mode'', we plan on ensuring compatibility of this mode with mobile interfaces, such as tablets and phones.

We would also like to explore expanding the Sanguine annotation system, to better address the data integrity issue described in \textbf{Discussion and Limitation}. A system that allows analysts to annotate on any components would enhance our current text-based chart annotation structure, and further facilitate communication between PBM experts and clinicians.

Even though we focus on PBM data in this design study, the techniques we used can likely be applied to other areas of medicine, where analysts want to quickly view outcomes of interest and their relationships with variables of interest.

\section*{Acknowledgements}
We thank Dr.\ Vikas Sharma, Dr.\ Joseph Tonna, Dr.\ Josh Zimmerman, and Dr.\ Candice Morrisey for their input and feedback. We also thank the support from the Enterprise Data Warehouse and the Center for High Performance Computing at the University of Utah. 

\section*{Funding}

This project is funded by ARUP laboratories; the National Science Foundation (IIS 1751238); and the National Institutes of Health (1S10OD021644-01A1).

\bibliographystyle{unsrt}  

\bibliography{references}  

\begin{thebibliography}{10}

\bibitem{pfuntner_most_2013}
Anne Pfuntner, Lauren~M. Wier, and Carol Stocks.
\newblock Most {Frequent} {Procedures} {Performed} in {U}.{S}. {Hospitals},
  2010: {Statistical} {Brief} \#149.
\newblock In {\em Healthcare {Cost} and {Utilization} {Project} ({HCUP})
  {Statistical} {Briefs}}. Agency for Healthcare Research and Quality (US),
  Rockville (MD), 2013.

\bibitem{ellingson_continued_2017}
Katherine~D. Ellingson, Mathew R.~P. Sapiano, Kathryn~A. Haass, Alexandra~A.
  Savinkina, Misha~L. Baker, Koo-Whang Chung, Richard~A. Henry, James~J.
  Berger, Matthew~J. Kuehnert, and Sridhar~V. Basavaraju.
\newblock Continued decline in blood collection and transfusion in the {United}
  {States}–2015.
\newblock {\em Transfusion}, 57(Suppl 2):1588--1598, June 2017.

\bibitem{the_joint_commission_and_the_american_medical_association-_convened_physician_consortium_for_performance_improvement_proceedings_2012}
The Joint Commission~{and} the American Medical Association-Convened Physician
  Consortium~for Performance~Improvement.
\newblock Proceedings from the {National} {Summit} on {Overuse}, September
  2012.

\bibitem{delaney_transfusion_2016}
Meghan Delaney, Silvano Wendel, Rachel~S Bercovitz, Joan Cid, Claudia Cohn,
  Nancy~M Dunbar, Torunn~O Apelseth, Mark Popovsky, Simon~J Stanworth, Alan
  Tinmouth, Leo Van De~Watering, Jonathan~H Waters, Mark Yazer, and Alyssa
  Ziman.
\newblock Transfusion reactions: prevention, diagnosis, and treatment.
\newblock {\em The Lancet}, 388(10061):2825--2836, December 2016.

\bibitem{gross_patient_2015}
Irwin Gross, Burkhardt Seifert, Axel Hofmann, and Donat~R. Spahn.
\newblock Patient blood management in cardiac surgery results in fewer
  transfusions and better outcome.
\newblock {\em Transfusion}, 55(5):1075--1081, 2015.
\newblock \_eprint: https://onlinelibrary.wiley.com/doi/pdf/10.1111/trf.12946.

\bibitem{goodnough_patient_2012}
Lawrence~Tim Goodnough and Aryeh Shander.
\newblock Patient blood management.
\newblock {\em Anesthesiology}, 116(6):1367--1376, June 2012.

\bibitem{shander_bloodless_2012}
Aryeh Shander, Mazyar Javidroozi, Seth Perelman, Thomas Puzio, and Gregg Lobel.
\newblock From {Bloodless} {Surgery} to {Patient} {Blood} {Management}.
\newblock {\em Mount Sinai Journal of Medicine: A Journal of Translational and
  Personalized Medicine}, 79(1):56--65, January 2012.

\bibitem{frank_novel_2013}
Steven~M. Frank, Linda~M.S. Resar, James~A. Rothschild, Elizabeth~A. Dackiw,
  Will~J. Savage, and Paul~M. Ness.
\newblock A novel method of data analysis for utilization of red blood cell
  transfusion: {Analysis} of {RBC} {Utilization}.
\newblock {\em Transfusion}, 53(12):3052--3059, December 2013.

\bibitem{frank_variability_2012}
Steven~M. Frank, Will~J. Savage, Jim~A. Rothschild, Richard~J. Rivers, Paul~M.
  Ness, Sharon~L. Paul, and John~A. Ulatowski.
\newblock Variability in {Blood} and {Blood} {Component} {Utilization} as
  {Assessed} by an {Anesthesia} {Information} {Management} {System}:.
\newblock {\em Anesthesiology}, 117(1):99--106, July 2012.

\bibitem{hannon_waste_2015}
Timothy Hannon.
\newblock Waste {Not}, {Want} {Not}.
\newblock {\em American Journal of Clinical Pathology}, 143(3):318--319, March
  2015.

\bibitem{frank_reducing_2014}
Steven~M. Frank, Michael~J. Oleyar, Paul~M. Ness, and Aaron A.~R. Tobian.
\newblock Reducing {Unnecessary} {Preoperative} {Blood} {Orders} and {Costs} by
  {Implementing} an {Updated} {Institution}-specific {Maximum} {Surgical}
  {Blood} {Order} {Schedule} and a {Remote} {Electronic} {Blood} {Release}
  {System}:.
\newblock {\em Anesthesiology}, 121(3):501--509, September 2014.

\bibitem{metcalf_data-driven_2018}
R.~A. Metcalf, M.~B. Pagano, J.~R. Hess, J.~Reyes, J.~D. Perkins, and M.~I.
  Montenovo.
\newblock A data-driven patient blood management strategy in liver
  transplantation.
\newblock {\em Vox Sanguinis}, May 2018.

\bibitem{kerzner_framework_2019}
Ethan Kerzner, Sarah Goodwin, Jason Dykes, Sara Jones, and Miriah Meyer.
\newblock A {Framework} for {Creative} {Visualization}-{Opportunities}
  {Workshops}.
\newblock {\em IEEE Transactions on Visualization and Computer Graphics},
  25(1):748--758, January 2019.

\bibitem{brehmer_multi-level_2013}
Matthew Brehmer and Tamara Munzner.
\newblock A {Multi}-{Level} {Typology} of {Abstract} {Visualization} {Tasks}.
\newblock {\em IEEE Transactions on Visualization and Computer Graphics},
  19(12):2376--2385, December 2013.

\bibitem{gleicher_considerations_2018}
M.~Gleicher.
\newblock Considerations for {Visualizing} {Comparison}.
\newblock {\em IEEE Transactions on Visualization and Computer Graphics},
  24(1):413--423, 2018.

\bibitem{rind_interactive_2013}
Alexander Rind.
\newblock Interactive {Information} {Visualization} to {Explore} and {Query}
  {Electronic} {Health} {Records}.
\newblock {\em Foundations and Trends® in Human–Computer Interaction},
  5(3):207--298, 2013.

\bibitem{caban_visual_2015}
J.~J. Caban and D.~Gotz.
\newblock Visual analytics in healthcare - opportunities and research
  challenges.
\newblock {\em Journal of the American Medical Informatics Association},
  22(2):260--262, March 2015.

\bibitem{shneiderman_improving_2013}
Ben Shneiderman, Catherine Plaisant, and Bradford~W. Hesse.
\newblock Improving {Healthcare} with {Interactive} {Visualization}.
\newblock {\em Computer}, 46(5):58--66, May 2013.

\bibitem{west_innovative_2015}
Vivian~L. West, David Borland, and W.~Ed Hammond.
\newblock Innovative information visualization of electronic health record
  data: a systematic review.
\newblock {\em Journal of the American Medical Informatics Association},
  22(2):330--339, March 2015.
\newblock Publisher: Oxford Academic.

\bibitem{plaisant_lifelines_2003}
Catherine Plaisant, Richard Mushlin, Aaron Snyder, Jia Li, Dan Heller, and Ben
  Shneiderman.
\newblock {LifeLines}: {Using} {Visualization} to {Enhance} {Navigation} and
  {Analysis} of {Patient} {Records}.
\newblock In BENJAMIN~B. Bederson and BEN Shneiderman, editors, {\em The
  {Craft} of {Information} {Visualization}}, Interactive {Technologies}, pages
  308--312. Morgan Kaufmann, San Francisco, January 2003.

\bibitem{bui_timeline_2007}
Alex A.~T. Bui, Denise~R. Aberle, and Hooshang Kangarloo.
\newblock {TimeLine}: {Visualizing} {Integrated} {Patient} {Records}.
\newblock {\em IEEE Transactions on Information Technology in Biomedicine},
  11(4):462--473, July 2007.
\newblock Conference Name: IEEE Transactions on Information Technology in
  Biomedicine.

\bibitem{faiola_medical_2014}
Anthony Faiola.
\newblock Medical information visualization assistant system and method,
  February 2014.

\bibitem{gotz_decisionflow_2014}
David Gotz and Harry Stavropoulos.
\newblock {DecisionFlow}: {Visual} {Analytics} for {High}-{Dimensional}
  {Temporal} {Event} {Sequence} {Data}.
\newblock {\em IEEE Transactions on Visualization and Computer Graphics},
  20(12):1783--1792, December 2014.

\bibitem{wongsuphasawat_exploring_2012}
K.~Wongsuphasawat and D.~Gotz.
\newblock Exploring {Flow}, {Factors}, and {Outcomes} of {Temporal} {Event}
  {Sequences} with the {Outflow} {Visualization}.
\newblock {\em IEEE Transactions on Visualization and Computer Graphics},
  18(12):2659--2668, December 2012.

\bibitem{krause_supporting_2016}
Josua Krause, Adam Perer, and Harry Stavropoulos.
\newblock Supporting {Iterative} {Cohort} {Construction} with {Visual}
  {Temporal} {Queries}.
\newblock {\em IEEE Transactions on Visualization and Computer Graphics},
  22(1):91--100, January 2016.
\newblock Conference Name: IEEE Transactions on Visualization and Computer
  Graphics.

\bibitem{rogers_composervisual_2019}
Jen Rogers, Nicholas Spina, Ashley Neese, Rachel Hess, Darrel Brodke, and
  Alexander Lex.
\newblock Composer—{Visual} {Cohort} {Analysis} of {Patient} {Outcomes}.
\newblock {\em Applied Clinical Informatics}, 10(02):278--285, March 2019.

\bibitem{bernard_visual-interactive_2015}
Jürgen Bernard, David Sessler, Thorsten May, Thorsten Schlomm, Dirk Pehrke,
  and Jörn Kohlhammer.
\newblock A {Visual}-{Interactive} {System} for {Prostate} {Cancer} {Cohort}
  {Analysis}.
\newblock {\em IEEE Computer Graphics and Applications}, 35(3):44--55, May
  2015.
\newblock Conference Name: IEEE Computer Graphics and Applications.

\bibitem{glueck_phenostacks_2017}
Michael Glueck, Alina Gvozdik, Fanny Chevalier, Azam Khan, Michael Brudno, and
  Daniel Wigdor.
\newblock {PhenoStacks}: {Cross}-{Sectional} {Cohort} {Phenotype} {Comparison}
  {Visualizations}.
\newblock {\em IEEE Transactions on Visualization and Computer Graphics},
  23(1):191--200, January 2017.
\newblock Conference Name: IEEE Transactions on Visualization and Computer
  Graphics.

\bibitem{bernard_using_2018}
Jürgen Bernard, David Sessler, Joern Kohlhammer, and Roy~A Ruddle.
\newblock Using dashboard networks to visualize multiple patient histories: a
  design study on post-operative prostate cancer.
\newblock {\em IEEE transactions on visualization and computer graphics},
  25(3):1615--1628, 2018.
\newblock Publisher: IEEE.

\bibitem{lee_web-based_2016}
Joon Lee, Evan Ribey, and James~R. Wallace.
\newblock A web-based data visualization tool for the {MIMIC}-{II} database.
\newblock {\em BMC Medical Informatics and Decision Making}, 16(1):15, February
  2016.

\bibitem{rajwan_medical_2010}
Yair~G. Rajwan and George~R. Kim.
\newblock Medical information visualization conceptual model for
  patient-physician health communication.
\newblock In {\em Proceedings of the 1st {ACM} {International} {Health}
  {Informatics} {Symposium}}, {IHI} '10, pages 512--516, New York, NY, USA,
  November 2010. Association for Computing Machinery.

\bibitem{hakone_proact_2017}
Anzu Hakone, Lane Harrison, Alvitta Ottley, Nathan Winters, Caitlin Gutheil,
  Paul K.~J. Han, and Remco Chang.
\newblock {PROACT}: {Iterative} {Design} of a {Patient}-{Centered}
  {Visualization} for {Effective} {Prostate} {Cancer} {Health} {Risk}
  {Communication}.
\newblock {\em IEEE Transactions on Visualization and Computer Graphics},
  23(1):601--610, January 2017.
\newblock Conference Name: IEEE Transactions on Visualization and Computer
  Graphics.

\bibitem{galvez_visual_2014}
Jorge~A Gálvez, Luis Ahumada, Allan~F Simpao, Elaina~E Lin, Christopher~P
  Bonafide, Dhruv Choudhry, William~R England, Abbas~F Jawad, David Friedman,
  Debora~A Sesok-Pizzini, and Mohamed~A Rehman.
\newblock Visual analytical tool for evaluation of 10-year perioperative
  transfusion practice at a children's hospital.
\newblock {\em Journal of the American Medical Informatics Association},
  21(3):529--534, May 2014.

\bibitem{gratzl_lineup_2013}
Samuel Gratzl, Alexander Lex, Nils Gehlenborg, Hanspeter Pfister, and Marc
  Streit.
\newblock {LineUp}: {Visual} {Analysis} of {Multi}-{Attribute} {Rankings}.
\newblock {\em IEEE Transactions on Visualization and Computer Graphics
  (InfoVis '13)}, 19(12):2277--2286, 2013.

\bibitem{furmanova_taggle_2020}
Katarina Furmanova, Samuel Gratzl, Holger Stitz, Thomas Zichner, Miroslava
  Jaresova, Alexander Lex, and Marc Streit.
\newblock Taggle: {Combining} overview and details in tabular data
  visualizations.
\newblock {\em Information Visualization}, 19(2):114--136, 2020.
\newblock Publisher: SAGE Publications.

\bibitem{cutler_trrack_2020}
Zachary~T Cutler, Kiran Gadhave, and Alexander Lex.
\newblock Trrack: {A} {Library} for {Provenance} {Tracking} in {Web}-{Based}
  {Visualizations}.
\newblock preprint, Open Science Framework, July 2020.

\bibitem{brown_into_2011}
Barry Brown, Stuart Reeves, and Scott Sherwood.
\newblock Into the wild: challenges and opportunities for field trial methods.
\newblock In {\em Proceedings of the 2011 annual conference on {Human} factors
  in computing systems - {CHI} '11}, page 1657, Vancouver, BC, Canada, 2011.
  ACM Press.

\bibitem{metcalf_association_2019}
Ryan~A. Metcalf, Sandra~K. White, Scott Potter, Reed Barney, Cheri Hunter,
  Michael White, Toby Enniss, Charles Galaviz, Santosh Reddy, Nathan Wanner,
  Robert~L. Schmidt, and Robert Blaylock.
\newblock The association of inpatient blood utilization and diagnosis-related
  group weight: implications for risk-adjusted benchmarking.
\newblock {\em Transfusion}, 59(7):2316--2323, 2019.
\newblock \_eprint: https://onlinelibrary.wiley.com/doi/pdf/10.1111/trf.15343.

\bibitem{mccurdy_framework_2019}
Nina Mccurdy, Julie Gerdes, and Miriah Meyer.
\newblock A {Framework} for {Externalizing} {Implicit} {Error} {Using}
  {Visualization}.
\newblock {\em IEEE Transactions on Visualization and Computer Graphics},
  25(1):925--935, January 2019.

\bibitem{sedlmair_design_2012}
Michael Sedlmair, Miriah Meyer, and Tamara Munzner.
\newblock Design {Study} {Methodology}: {Reflections} from the {Trenches} and
  the {Stacks}.
\newblock {\em IEEE Transactions on Visualization and Computer Graphics},
  18(12):2431--2440, December 2012.

\bibitem{rogers_insights_2021}
J.~Rogers, A.~H. Patton, L.~Harmon, A.~Lex, and M.~Meyer.
\newblock Insights {From} {Experiments} {With} {Rigor} in an {EvoBio} {Design}
  {Study}.
\newblock {\em IEEE Transactions on Visualization and Computer Graphics},
  27(2):1106--1116, February 2021.
\newblock Conference Name: IEEE Transactions on Visualization and Computer
  Graphics.

\end{thebibliography}

\end{document}